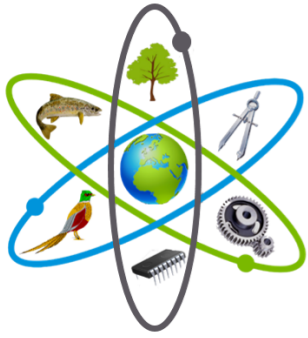

- *RESEARCH ARTICLE*-

# Effect of Window Size for Detection of Abnormalities in Respiratory Sounds

Osman Balli [1], Yakup Kutlu[2]

[1]Department of Computer Engineering, Iskenderun Technical University, Turkey
[2]Department of Computer Engineering, Iskenderun Technical University, Turkey

**Abstract**

The recording of respiratory sounds was of significant benefit in the diagnosis of abnormalities in respiratory sounds. The duration of the sounds used in the diagnosis affects the speed of the diagnosis. In this study, the effect of window size on diagnosis of abnormalities in respiratory sounds and the most efficient recording time for diagnosis were analyzed. First, window size was applied to each sound in the data set consisting of normal and abnormal breathing sounds, 0.5 second and from 1 to 20 seconds Increased by 1 second. Then, the data applied to window size was inferred feature extraction with Mel Frequency Cepstral Coefficient (MFCC) and the performance of each window was calculated using the leave one-out classifier and the k-nearest neighbor (KNN) algorithm. As a result, it was determined that the data was significant with an average performance of 92.06% in the records between 2 and 10 seconds.

## Introduction

Listening to respiratory sounds directly or indirectly is called auscultation. Diagnosis can be simple or difficult depending on the condition of the disease in examinations performed by auscultation. This situation is proportional to the experience of the physician. With the developing technology, new devices are being produced to make the auscultation process easier and more reliable. Electronic auscultation is called the technique of using an electronic stethoscope or microphone during auscultation process. Respiratory sounds contain very important information about the physiology and pathology of lung and respiratory obstruction. Differences between Normal respiratory (lung) sounds and abnormal ones (e.g. crackling and wheezing) are crucial to making a definitive medical diagnosis (Sovijarvi, Malmberg, Charbonneau, & Vandershoot, 2000). Thanks to the electronic auscultation the breathing sounds heard in the auscultation can be recorded during the process. These recorded sounds are also used in education or in automatic recognition of respiratory sounds.

With the recording of respiratory sounds, it became quite easy to analyze the differences. In addition to being able to distinguish between Normal and abnormal sounds, recordings can be used to examine the sounds in greater detail and make precise diagnoses. With the advancement of technology and the impact of artificial intelligence on every area, many studies are being done for automatic recognition of breathing sounds. Automatic recognition time is short and reliable is one of the important issues of these studies.

In the literature, there are different methods in the extraction and classification of features from respiratory sounds (Altan & Kutlu, 2016). Deep Extreme Learning Machines, which is seen as some basic biological learning structures, was used for Chronic Obstructive Pulmonary Disease (COPD) and obtained a high accuracy of 91.39% (Altan, Kutlu, & Yayık, 2018). In recent years, deep learning, a highly popular machine learning approach have also been tried COPD and asthma which are major diseases that can be diagnosed from respiratory sounds and achieved highly successful results (Allahverdi, Altan, & Kutlu, 2018) (Altan & Kutlu, 2018). One of the most commonly used methods for automatic recognition of respiratory sounds is the Mel Frequency Cepstrum Coefficient (MFCC) technique (Palaniappan, Sundaraj, & Sundaraj, 2014a). Studies for automatic recognition of respiratory sounds: Using a combination of MFCC and Gaussian Mixture Models (GMM), Bahoura (Bahoura, 2009) compared it to other methods. Palanippan et al. (Palaniappan, Sundaraj, & Sundaraj, 2014b) calculated MFCC coefficients and used Support Vector Machines (SVM) and k nearest neighbor (KNN) algorithms. In the study of Aras and Gangal, Hjords parameters, mean, standard deviation, skewness, kurtosis and entropy values were obtained from MFCC and tested using SVM, KNN, LDA, NBA algorithms (Aras & Gangal, 2017). In another study, cracks and friction rubs were tried to be classified, in which the minimum, maximum, average and standard deviation of the MFCC were calculated (Khan & Ahmed, 2016).

Window size analysis of respiratory sounds provides information about when the sound is more distinctive. The most efficient recording time determined by Window size can increase the efficiency of the system by using automatic diagnostic systems. In this study, the records in the data set were divided into normal and abnormal. All records were first applied to window size. The minimum, maximum, average, standard deviation, kurtosis and skewness values of each size were calculated using MFCC from the records applied to Window size. The data were classified with the Leave one-Out Cross Validation method and their accuracy was calculated with the KNN algorithm. Each size's accuracy was compared and the most efficient time to recognize respiratory sounds was analyzed.

**Database**

In this study, Rocha et al. his respiratory sounds data set (Rocha, et al., 2017) was used. Sounds are divided into normal (healthy) and abnormal (asthma, Bronchiectasis, Bronchiolitis, Pneumonia, lower respiratory tract infection (LRTI), Upper Respiratory Tract Infection (URTI)). A total of 127 sound recordings were used, including 35 normal and 92 abnormal. Each of the data is 20 seconds and the sampling frequency is 44100 Hz.

**Window Size**

Sounds can show more distinctive features at different time intervals. Based on the times when there are more distinguishing features, studies on Sounds will speed up. Detecting these characteristics in respiratory sounds is crucial for automatic diagnosis of abnormalities. It can shorten the duration of automatic diagnosis and also contribute to the creation of a more reliable system.

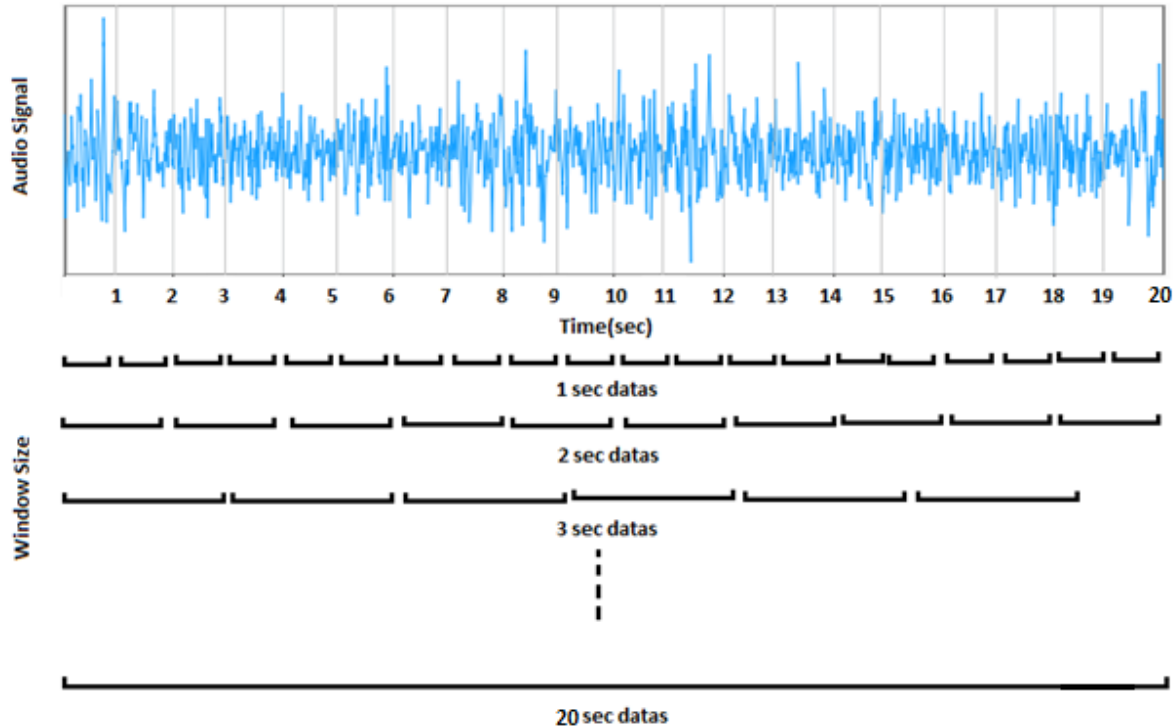

Figure 1. Window Size in Implementation

In this study, window size was applied to each sound in the data set consisting of normal and abnormal breathing sounds, 0.5 second and from 1 to 20 seconds Increased by 1 second. With the data set applied to Window size, character analysis of sounds at different times was performed and meaningful results were obtained.

**Mel Frequency Cepstrum Coefficients based Feature Extruction**

Mel Frequency Cepstrum Coefficient (MFCC) is a method of demonstrating the spectral characteristics of signals. It is extremely effective and popular for feature extraction in speech signals, especially in recent periods. MFCC has also been used in respiratory sounds in general and significant accuracies have been achieved (Aras & Gangal, 2017). Figure 2 shows the processes used to obtain of MFCC step by step.

*Natural and Engineering Sciences*

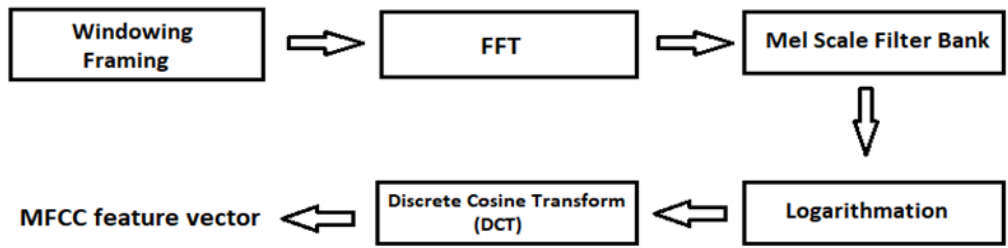

Figure 2. Block Diagram of MFCC

The respiratory sound signal devided into frames overlapping each order by %50. Then the triangular bandpass filter known as mel scale filter bank was applied to the coefficients obtained by applying Fast Fourier Transform (FFT). Using (1), the real frequency scale obtained from this converted to the Mel logarithmic scale;

$$Mel(f) = 2595 \log(1 + \frac{f}{700})$$
(1)

The logarithmic scale is then converted to time through the use of a discrete cosine transform, and the output is the set of MFCCs. The MFCCs obtained from the respiratory sounds are used as features in like the KNN classifiers. In this study, 14 MFCCs were extracted for the classification of the respiratory sounds. Using MFCC coefficients, 6 features (minimum, maximum, kurtosis, skewness, standard deviation, mean) were derived. Z-Score, which is one of the standardization methods that allows the units in the data set to be stacked into a common range of units, was used in these properties. Z-score of applied properties has been used in classification.

**Classification and Training**

The K Nearest Neighbor (KNN) algorithm used in this study is one of the simple learning methods that does not require the probability distribution of classes. The distance of each data in the test set to each data in the training set is calculated. For each data in the test set, the nearest K sample is selected from the training set. The test data is assigned to the class in which the selected K sample contains the maximum. During the training phase, Leave-One-Out Cross-Validation (LOOCV) method was used to make the best use of the training data and eliminate the problems of random selection (Aydemir & Kayıkçıoğlu, 2011). This method allows each data to be used as a test. The LOOCV method and KNN algorithm are preferred in this study because they yield successful results (Aras & Gangal, 2017).

**Results and Discussion**

As a result of this study, breathing sounds were found to have more pronounced characteristics at different times. It was found that after a certain period of time, the increase in the properties of sound made no sense to the data and reduced performance. In Figure 3, the data was more significant with an average accuracy of 92.06% between 2 and 10 seconds. In the remaining times the data has lost meaning, especially in there was a significant decrease after the 10.second. My highest accuracy is 93.21% with seen in 5.second. In future studies, examination of the first 10 seconds of respiratory sounds will provide more meaningful results. In this study, it was analyzed that 5-second sound recordings were sufficient for automatic diagnosis of respiratory sounds and high results could be obtained.

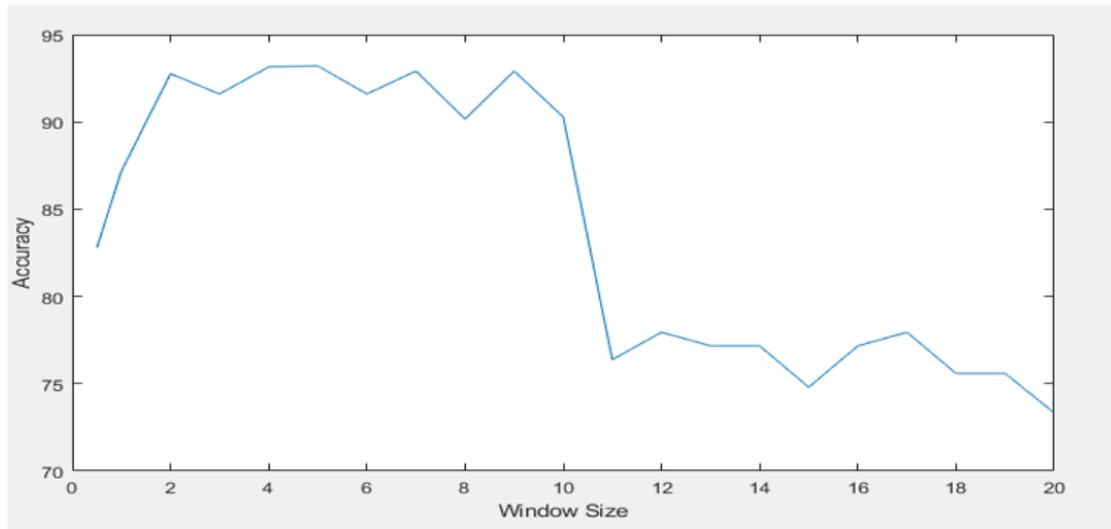

Figure 3. Accuracy versus window size

**Note:** This paper is presented in the International Conference on Artificial Intelligence towards Industry 4.0 held on November 14-16, 2019 at Iskenderun Technical University, Iskenderun, Turkey.